\documentclass[seceq]{ptptex}

\usepackage{amsmath,amsfonts,latexsym,amssymb} 
\usepackage{verbatim,amsthm,graphicx} 

\def\K{{\cal K}}

\def\N{{\cal N}}

\font\tenscr=rsfs10 scaled1100
\font\sevenscr=rsfs7 
\font\fivescr=rsfs5 
\skewchar\tenscr='177 
\skewchar\sevenscr='177
\skewchar\fivescr='177 
\newfam\scrfam 
\textfont\scrfam=\tenscr
\scriptfont\scrfam=\sevenscr 
\scriptscriptfont\scrfam=\fivescr

\def\linebreak{\hfill\break}

\def\bra<#1|{\langle #1\rvert}
\def\ket|#1>{\lvert#1 \rangle}
\def\braket<#1|#2>{\langle #1|#2 \rangle}

\def\pfrac#1#2{\left(\frac{#1}{#2}\right)}
\def\const{\text{const}}
\def\for{\text{for}}

\def\tend{\rightarrow}

\def\equivalent{\quad\Leftrightarrow\quad}
\def\therefore{\mbox{\setbox0=\hbox{X}\hbox{$\ldotp$}\raise0.7\ht0\hbox{$\ldotp$}\hbox{$\ldotp$}} \quad }
\def\because{\mbox{\setbox0=\hbox{X}\raise0.7\ht0\hbox{$\ldotp$}\hbox{$\ldotp$}\raise0.7\ht0\hbox{$\ldotp$}}\kern0pt }
\def\for{\text{\ for\ }}
\def\r#1{{\rm #1}}

\def\bm#1{\boldsymbol{#1}}

\def\SO{{\rm SO}}

\def\Frac(#1/#2){\left(\frac{#1}{#2}\right)}

\def\Order#1{\r{O}\!\left(#1\right)}

\def\inpare#1{\left(#1\right)}
\def\inrbra#1{\left\{ #1 \right\}}
\def\insbra#1{\left[ #1 \right]}

\def\Lie{\hbox{\rlap{$\cal L$}$-$}}

\def\Eq#1{\begin{equation} #1 \end{equation}}

\def\Eqr#1{\begin{eqnarray} #1 \end{eqnarray}}

\def\Eqrsub#1{\begin{subequations}\Eqr{#1}\end{subequations}}
\def\Eqrsubl#1#2{\begin{subequations}\label{#1}\Eqr{#2}\end{subequations}}

\def\Bitm{\begin{itemize}}
\def\Eitm{\end{itemize}}
\def\Blist#1#2{\begin{list}{#1}{\parsep=0pt \itemsep=0pt%
  \listparindent=0pt #2}}
\def\Elist{\end{list}}
\long\def\ignore#1#2{\def\ignoreflag{#1}\long\def\tmptext{#2}
  \ifnum\ignoreflag>1 #2 \fi}

\theoremstyle{definition}

\newtheorem{theorem}{Theorem}[]
\newtheorem{proposition}{Proposition}[]
\def\THB{{\mathbb T}}
\def\VHB{{\mathbb V}}
\def\SHB{{\mathbb S}}

\title{
Perturbative Uniqueness of Black Holes near the Static Limit \\
in All Dimensions
}

\author{Hideo {\sc Kodama}%
\footnote{E-mail: kodama@yukawa.kyoto-u.ac.jp} 
}

\inst{Yukawa Institute for Theoretical Physics, Kyoto University \\
Kyoto  606-8502, Japan 
}

\abst{
The behaviour of stationary gravitational perturbations is studied 
for generalised static black holes in spacetimes of greater than 
three dimensions, using the formulation developed by the present 
author and Ishibashi. For the case in which the horizon has a 
spatial section with constant curvature, it is proved that 
irrespective of the value of the cosmological constant $\Lambda$, 
there exists no stationary perturbation that is regular at the 
horizon(s) and falls off at infinity in the case $\Lambda\le0$, 
except for those corresponding to the stationary rotation of black 
holes and the variation of the background parameters. This result 
indicates that regular neutral black hole solutions that are either 
asymptotically flat, de Sitter or anti-de Sitter can be parametrised 
by mass, (multiple component) angular momentum and $\Lambda$ near 
the spherically symmetric and static limit. A similar conclusion is 
obtained for topological black holes. It is also pointed out that 
this perturbative uniqueness near the static limit may not hold in 
the case in which the horizon geometry is described by a generic 
Einstein space with non-constant sectional curvature. Further, 
non-uniqueness in the asymptotically anti-de Sitter case under a 
weaker boundary condition at infinity related to the AdS/CFT 
argument is discussed.}

\begin{document}

\maketitle

\section{Introduction}

In recent years, inspired by proposals of various higher-dimensional 
universe models\cite{Horava.P&Witten1996,Horava.P&Witten1996a,%
Lukas.A&&1999,Lukas.A&&1999a,Lukas.A&Ovrut&Waldram1999,%
Arkani-Hamed.N&Dimopoulos&Dvali1998,%
Antoniadis.I&&1998,Randall.L&Sundrum1999a,Randall.L&Sundrum1999b}, 
the investigation of black holes in higher dimensions has become 
quite active. One of the main issues in these investigations is 
whether the celebrated black hole uniqueness theorem in four 
dimensions%
\cite{Heusler.M1996B,Heusler.M1998,Chrusciel.P1996} also holds in 
higher dimensions.

Results obtained to this time concerning this issue  are quite 
intriguing\cite{Kodama.H2004A}. On one hand, it has been shown that 
Israel's theorem on the rigidity and uniqueness of static black 
holes also holds in higher dimensions, that is, the 
asymptotically flat and static regular black hole solutions are 
always spherically symmetric for the vacuum system%
\cite{Hwang.S1998,Gibbons.G&Ida&Shiromizu2003}. 
This rigidity theorem has been extended to the electro-vacuum and  
Einstein-Maxwell-Dilaton systems with non-degenerate 
horizons%
\cite{Gibbons.G&Ida&Shiromizu2002a,Gibbons.G&Ida&Shiromizu2002b}. 

On the other hand, concerning rotating black holes, we now have two 
families of asymptotically flat regular vacuum solutions with 
different horizon topologies in five dimensions, the Myers-Perry 
solution\cite{Myers.R&Perry1986} and the Emparan-Reall black ring 
solution\cite{Emparan.R&Reall2002a}. The former is a 
higher-dimensional analogue of the Kerr solution, and its black hole 
surface is diffeomorphic to a sphere, while the latter has no 
analogue in four dimensions and its horizon has a spatial section 
diffeomorphic to $S^2\times S^1$. Thus, the uniqueness theorem is 
violated for rotating black holes. Recently, it has also been shown 
that a new family of black ring solutions violating the uniqueness 
appear when the system is coupled with form 
fields\cite{Emparan.R2004A}, although the uniqueness holds for 
supersymmetric black holes in the 
five-dimensional minimal $N=1$ supergravity 
theory\cite{Reall.H2003}. Further, Emparan and Myers have pointed 
out that the Myers-Perry solutions in greater than five dimensions 
may suffer from a Gregory-Laflamme type instability for high angular 
momenta and argued that there may exist a new family of rotating 
solutions that bifurcates from the Myers-Perry 
family maintaining the symmetry\cite{Emparan.R&Myers2003}. Such a symmetry preserving 
branch does not exist in five dimensions, as shown by Morisawa and 
Ida\cite{Morisawa.Y&Ida2004A}.
 
Along with these results in the asymptotically flat case, 
non-uniqueness has also been observed in Kaluza-Klein-type 
spacetimes. For example, Kudoh and Wiseman numerically constructed a 
family of non-uniform black string solutions in six 
dimensions\cite{Wiseman.T2003,Kudoh.H&Wiseman2004,Kudoh.H2004}. This 
family bifurcates from the uniform black string solution at the 
Gregory-Laflamme critical point. As discussed by Horowitz and 
Maeda\cite{Horowitz.G&Maeda2001}, this family cannot be smoothly 
connected to a family of localised black hole solutions with 
spherical horizon. In fact, Wiseman numerically found that these two 
families bifurcate at a singular solution with a conic 
horizon\cite{Wiseman.T2003a}, confirming Kol's 
conjecture\cite{Kol.B&Wiseman2003}.

One common feature of these new families of solutions giving rise to 
non-uniqueness is that they do not have a static and spherically 
symmetric limit. This feature, together with the uniqueness of 
asymptotically flat static black holes, strongly indicates that the 
violation of uniqueness in higher dimensions occurs only when the 
black horizon is significantly deformed due to high angular 
momentums or has a non-trivial topology. The main purpose of the 
present paper is to confirm this observation by searching for 
regular stationary perturbations of static and maximally symmetric 
black hole 
solutions. To be precise, we prove the following theorem. 

\medskip
\begin{theorem} \em
For any spherically symmetric vacuum solution that represents a 
regular black hole spacetime of dimension $d (\ge4)$, a scalar-type 
perturbation corresponding to a variation of the black hole mass and 
vector-type perturbations representing rotation of the black hole 
are the only stationary bounded perturbations that are regular at 
and outside the horizon (and fall off at infinity in the case  
$\Lambda\le0$). These exceptional vector perturbations can be 
parametrised by $[(d-1)/2]$ parameters after identification with  
background isometries, irrespective of the value of the 
cosmological constant $\Lambda$, where $[(d-1)/2]$ is the rank of 
$\SO(d-1)$. A similar result holds for a regular static topological 
black hole spacetime for which a spatial section of the horizon is 
an $n$-dimensional compact space with non-positive constant 
curvature $K$. The only difference between the two cases is the 
number of degrees of freedom of the exceptional vector perturbation, 
which is $d-2$ for $K=0$ and zero for $K<0$. 
\end{theorem} 
\medskip

For the asymptotically constant-curvature vacuum case, this theorem 
implies that the Myers-Perry solution for $\Lambda=0$ and the 
Gibbons-L\"u-Page-Pope solution\cite{Gibbons.G&&2004A} for 
$\Lambda\not=0$ are the only regular black hole solutions near the 
static and spherically symmetric limit. Further, it shows that a 
similar perturbative uniqueness holds for topological black holes as 
well. Because no uniqueness theorem has been proved for 
asymptotically de Sitter black holes in four dimensions, this 
theorem also has a non-trivial implication for the four-dimensional 
case.

The present paper is organised as follows. In the next section, we 
give a brief summary of a gauge-invariant formulation for 
stationary perturbations of generalised static vacuum black holes. 
The basic equations given there are basically specialisations to the 
case of stationary perturbations of the equations for generic 
perturbations derived in Refs. 
\citen{Kodama.H&Ishibashi&Seto2000,Kodama.H&Ishibashi2003,%
Kodama.H&Ishibashi2003A,Kodama.H&Ishibashi2004}, but some are new. 
In order for solutions to these basic equations to represent regular 
perturbations of black hole spacetimes, 
an appropriate boundary condition at the horizon and an asymptotic 
condition at infinity should be satisfied. These boundary conditions 
are specified in \S3, and it is shown that the exceptional vector 
perturbations indeed do satisfy them. Then, in the subsequent three 
sections, it is proved that, except for the exceptional vector 
perturbations and the trivial scalar perturbations corresponding to 
variations of the background parameters, there exists no stationary 
solution satisfying the boundary conditions to the basic equations 
for the asymptotically flat, de Sitter and anti-de Sitter cases, 
respectively. Section 7 is devoted to concluding remarks. 

\section{Basic Equations for Stationary Perturbations}

Our analysis of black hole perturbations fully utilises the 
gauge-invariant formulation developed in Refs. 
\citen{Kodama.H&Ishibashi&Seto2000,Kodama.H&Ishibashi2003,%
Ishibashi.A&Kodama2003} and \citen{Kodama.H&Ishibashi2004}. In this 
section, we briefly summarise the basic concepts of that formulation 
and give master equations for stationary perturbations.  

\subsection{Background spacetime}

In the present paper, we consider the 
background spacetime whose metric has the form 
\Eq{
ds^2= g_{ab}(y)dy^a dy^b + r^2(y) d\sigma_n^2,
\label{GeneralisedStaticMetric}
}
where $g_{ab}(y)dy^a dy^b$ is the static metric of a two-dimensional 
spacetime $\N^2$, and $d\sigma_n^2=\gamma_{ij}(z)dz^i dz^j$ is the 
metric of an $n$-dimensional complete, compact Einstein space $\K^n$ 
whose Ricci curvature $\hat R_{ij}$ satisfies the condition %
\Eq{
\hat R_{ij}=(n-1)K \gamma_{ij}
}
with $K=0,\pm1$. Although perturbative uniqueness can be proved only 
in the case in which $\K^n$ has a constant curvature, most analysis 
in the present paper is done without assuming this condition, in 
order to see what happens when $\K^n$ does not have a constant 
curvature. 

We assume that this metric satisfies the vacuum Einstein equations 
for $(n+2)$-dimensional spacetimes and represents a black hole. 
Then, the two-dimensional metric $g_{ab}$ is given by 
\Eq{
g_{ab}(y)dy^a dy^b =-f(r)dt^2+ \frac{dr^2}{f(r)},
\label{GeneralisedStaticMetric:2Dpart}
}
where 
\Eq{
f(r)=K-\pfrac{r_0}{r}^{n-1}-\lambda r^2;\quad
\lambda=\frac{2\Lambda}{n(n+1)}.
\label{GeneralisedStaticMetric:f}
}
This metric describes a regular black hole spacetime only when 
$\lambda$ satisfies the condition\cite{Kodama.H&Ishibashi2004}%
\footnote{For the interpretation of this solution in the case $K\le0$ as a topological black hole, see Refs. \citen{Aminneborg.S&&1996} and \citen{Birmingham.D1999}.} 
\Eq{
\lambda < \left\{ \begin{array}{ll}
       \lambda_{c2}:=\frac{n-1}{n+1}2^{\frac{n+1}{n-1}}r_0^{-2}
              &\for K=1,\\
       0  &\for K=0,-1.
       \end{array}\right.
}

Under this condition, in general, the region outside the black hole 
corresponds to the range $r_h<r<\infty$, where $r_h$ is the horizon 
radius, which is given by the (smallest positive) solution to   
\Eq{
f(r_h)=0\ \equivalent 
 \lambda=\frac{1}{r_h^2}\insbra{K-\pfrac{r_0}{r_h}^{n-1}}.
}
When we study the behaviour of perturbations, we consider this 
entire range for $\lambda\le0$. For $\lambda>0$, however, we  
consider only the region $r_h<r<r_c$, where $r_c$ is the radius of 
the cosmological horizon given by the larger positive root of 
$f(r)=0$, because the perturbative uniqueness of asymptotically de 
Sitter black holes can be proved only by looking at the behaviour of 
perturbations in this region.  

In this black hole background, the linearised Einstein equations can 
be reduced to simple master equations if we decompose 
the metric perturbation $\delta g_{\mu\nu}=h_{\mu\nu}$ into 
tensor-type, vector-type and scalar-type components, according to 
their transformation behaviour as tensors on 
$\K^n$\cite{Kodama.H&Ishibashi2003}. Now, we present these equations 
for the case of stationary perturbations.

\subsection{Tensor perturbations}

It is convenient to expand tensor perturbations in terms of the 
eigentensors satisfying
\Eq{
\hat \triangle_L \THB_{ij}=(k_T^2+2nK) \THB_{ij};\quad
\THB^i_i=0,\quad \hat D^j\THB_{ij}=0,
}
where $\hat \triangle_L$ is the Lichnerowicz operator
\Eq{
\hat \triangle_L h_{ij}=-\hat D\cdot\hat D h_{ij}-2\hat 
R_{ikjl}h^{kl}+2(n-1)K h_{ij}.
}
Here, $\hat D_i$ and $\hat R_{ijkl}$ represent the covariant 
derivative and the curvature tensor with respect to the metric 
$\gamma_{ij}$ on $\K^n$, respectively. When $\K^n$ is a constant 
curvature space, $\hat\triangle_L$ can be written simply as 
\Eq{
\hat \triangle_L=-\hat D\cdot\hat D +2nK.
}
Hence, $k_T^2$ corresponds to the eigenvalue of the operator $-\hat 
D\cdot\hat D$ on second-rank symmetric tensors. In this case, 
$k_T^2$ takes discrete and positive values, except in the case 
$K=0$. In particular, for $\K^n=S^n$, its spectrum is given by
\Eq{
k_T^2=l(l+n-1)-2,\quad l=2,3,\cdots .
}
In contrast, for $K=0$, i.e., when $\K^n$ is a torus $T^n$, the 
smallest eigenvalue is zero, and the corresponding eigentensors are 
given by trace-free constant symmetric matrices. There exist 
$n(n+1)/2-1$ such matrices, and these matrices represent variations 
of the moduli parameters of $T^n$.  

In general, the moduli degrees of freedom of a constant curvature 
space correspond to the tensor harmonics satisfying $k_T^2=-2K$. 
Hence, these constant matrices exhaust all moduli degrees of freedom 
for $K=0$, and there exists no moduli degree of freedom for $K=1$. 
Further, from the integral identity
\Eq{
2\int_{\K^n} d\Omega_n\, \hat D_{[i}\THB_{j]k} \hat 
D^{[i}\THB^{j]k}=(k_T^2+nK)\int_{\K^n}d\Omega_n\, 
\THB_{jk}\THB^{jk}\ge0,}
we obtain 
\Eq{
k_T^2 \ge -nK=n
}
for $K=-1$. Hence, for $K=-1$, there exist moduli degrees of freedom 
only when $n=2$, in accordance with Mostow's rigidity theorem%
\cite{Mostow.G1973B,Fujiwara.Y&Kodama&Ishihara1993,%
Barrow.J&Kodama2001,Barrow.J&Kodama2001a,Kodama.H2002}. Note that these tensor harmonics 
satisfying $k_T^2=-2K$ are the only non-trivial tensor harmonics for 
$n=2$.

Through the expansion in terms of these tensor harmonics, each mode 
of a stationary tensor perturbation can be expressed in terms of a 
gauge-invariant variable $H_T(r)$ as
\Eq{
h_{ab}=0,\ 
h_{ai}=0,\ 
h_{ij}=2r^2 H_T(r) \THB_{ij}.
}
Note that stationary tensor perturbations are always static. 

The linearised Einstein equations are reduced to%
\cite{Kodama.H&Ishibashi2003} 
\Eq{
\insbra{f (r^{n/2}H_T)'}'= r^{n/2-2} U_T H_T,
\label{MasterEq:Tensor:static}
}
where the prime denotes differentiation with respect to $r$, and 
\Eq{
U_T=\frac{n(n+2)}{4}f(r)
     +\frac{n(n+1)}{2}\pfrac{r_0}{r}^{n-1}
     +k_T^2 -(n-2)K.
}
Note that for the modes satisfying $k_T^2=-2K$, $H_T=\const$ is 
always a solution of this equation and represents a perturbation 
corresponding to the variation of the background metric with respect 
to its moduli parameters.

\subsection{Vector perturbations}
\label{subsec:BasicEqs:Vector}

Stationary vector perturbations can be expanded in terms of the 
vector harmonics satisfying
\Eq{
(\hat D\cdot \hat D+k_V^2)\VHB_i=0;\quad
\hat D_j \VHB^j=0
}
as
\Eq{
h_{ab}=0,\quad
h_{ai}=r f_a(r) \VHB_i,\quad
h_{ij}=2r^2 H_T(r) \VHB_{ij},
}
where
\Eq{
\VHB_{ij}:=-\frac{1}{2k_V}(\hat D_i \VHB_j + \hat D_j \VHB_i).
}
The eigenvalue $k_V^2$ takes discrete and positive values, except 
when $K=0$, and $k_V^2\ge n-1$ for 
$K=1$\cite{Kodama.H&Ishibashi2004}. In particular, for $\K^n=S^n$, 
the spectrum of $k_V^2$ is given by
\Eq{
k_V^2=l(l+n-1)-1,\quad l=1,2,\cdots.
}
In contrast, when $K=0$ and $\K^n$ is expressed as $\K^n=T^p\times 
\K'$, where $\K'$ is a Ricci flat space with no isometry, there 
exist $p$ independent covariantly constant vector fields providing 
harmonic vectors with $k_T^2=0$. These modes and those with 
$k_T^2=n-1$ for $K=1$ comprise the exceptional modes discussed 
below. 

First, for generic modes satisfying $k_V^2>(n-1)K$, the linearised 
Einstein equations can be written in terms of the gauge-invariant 
variables
\Eq{
F_a:= f_a + \frac{r}{k_V} D_a H_T
}
as\cite{Kodama.H&Ishibashi&Seto2000}
\Eqrsub{
&& D_a(r^{n+1}F^{(1)})-m_V r^{n-1}\epsilon_{ab} F^b=0,
\label{EinsteinEq:Vector:1}\\
&& D_a(r^{n-1}F^a)=0,
\label{EinsteinEq:Vector:2}
}
where
\Eqr{
&& F^{(1)}:=\epsilon^{ab}r D_a\pfrac{f_b}{r}
      =\epsilon^{ab}r D_a\pfrac{F_b}{r},\\
&& m_V:= k_V^2 -(n-1)K.
}
From the $a=t$ component of \eqref{EinsteinEq:Vector:1}, we find 
that $F^r=0$ for stationary perturbations, and 
\eqref{EinsteinEq:Vector:2} becomes trivial. Hence, the Einstein 
equations are equivalent to 
\Eqrsubl{EinsteinEq:Vector:static}{
&& \inpare{r^{n+1}F^{(1)}}'+m_V r^{n-1} F^t=0,
\label{EinsteinEq:Vector:static:1}\\
&& F^{(1)}=r\pfrac{F_t}{r}'.
}
Note that stationary vector perturbations are not static.

Next, for exceptional modes with $k_V^2=(n-1)K$, 
$\VHB_{ij}$ vanishes\cite{Kodama.H&Ishibashi2004}. (The factor 
$1/k_V$ in the definition of $\VHB_{ij}$ is introduced just for 
convenience and is not essential.) For these modes, $H_T$ does not 
appear, and perturbations are described by $f_a$ alone. Because 
$f_a$ transforms under the gauge transformation $\bar\delta 
z^i=L\VHB^i$ as $\bar\delta f_a=-rD_a L$, $f_r$ can be set to zero 
through such a gauge transformation. In this gauge, the Einstein 
equations are given by \eqref{EinsteinEq:Vector:static} with $m_V=0$ 
and $F^t=f^t$. This equation can be easily solved, and the general 
solution is given by
\Eq{
F^{(1)}=r\pfrac{f_t}{r}'=-\frac{J}{r^{n+1}}.
\label{ExpectionalMode:F1}
}
From this, we have
\Eq{
f_t=\frac{J}{n+1}\frac{1}{r^{n}}+Cr,
\label{ExceptionalMode:ft}
}
where $C$ is an integration constant. 

The gauge condition $f_r=0$ does not remove the gauge degree of 
freedom completely and leaves a residual gauge freedom such that 
$L=L(t)$. If we require that $f_t$ be independent of time, $L(t)$ is 
restricted to the form $L(t)=C' t$, where $C'$ is a constant. Under  
this gauge transformation, the above integration constant $C$ 
changes to $C-C'$. Hence, we can set this constant to zero through a 
gauge transformation. However, as discussed in the next section, 
this residual gauge degree of freedom plays an important role in 
proving the regularity of the above solution. In any case, this 
argument shows that the physical degrees of freedom of these 
exceptional modes can be parametrised by the single constant $J$ for 
each mode. As shown in Appendix \ref{Appendix:ExceptionalModes}, we 
can regard each of these modes as representing a rotational 
perturbation of a static black hole, and the parameter $J$ 
represents the total angular momentum of the perturbation.

Because the vector harmonics satisfying $k_V^2=(n-1)K$ are in 
one-to-one correspondence with the Killing vector fields of $\K^n$, 
these solutions form a linear space isomorphic to the linear space 
of Killing vector fields of $\K^n$. However, they are not all 
physically distinct, because two solutions related by an isometry of 
the background spacetime must be considered physically equivalent. 
In the case in which $\K^n=S^n$ and the orientation-preserving 
spatial isometry group is given by $\SO(n+1)$, the Killing vectors 
are in one-to-one correspondence with the antisymmetric matrices of 
rank $n+1$, and the transformation of a Killing vector by an 
isometry is mapped to a conjugate transformation of the 
corresponding antisymmetric matrix by an element of $\SO(n+1)$. 
Because the conjugate classes of these anti-symmetric matrices are 
classified according to their $[(n+1)/2]$ eigenvalues, where 
$[(n+1)/2]$ is the rank of $\SO(n+1)$, physically distinct 
exceptional modes are classified with respect to $[(n+1)/2]$  
constants. Thus, they have exactly the same number of degrees of 
freedom as that of the angular momentum parameters of the 
Myers-Perry solution\cite{Myers.R&Perry1986} and the 
Gibbons-L\"u-Page-Pope solution\cite{Gibbons.G&&2004A}. In fact, we 
can directly check that these exceptional modes can be obtained by 
expanding these solutions with respect to the angular momentum 
parameters. Note that when $\K^n$ is distinct from $S^n$, the 
parameter $J$ does not represent the angular momentum in the 
standard sense, although it is still related to a conserved quantity 
of the system. The 
number of physical degrees of freedom of the exceptional modes can 
differ from $[(n+1)/2]$ in that case.

\subsection{Scalar perturbations}
\label{subsec:BasicEqs:Scalar}

Scalar perturbations can be expanded in terms of harmonic 
functions satisfying
\Eq{
(\hat D\cdot\hat D + k^2)\SHB=0
}
and the vector and tensor harmonics derived from them,
\Eq{
\SHB_i=-\frac{1}{k} \hat D_i \SHB,\quad
\SHB_{ij}=\frac{1}{k^2} \hat D_i \hat D_j \SHB
          +\frac{1}{n}\gamma_{ij}\SHB.
}
Because we assume that $\K^n$ is compact, $k^2$ takes discrete 
values starting from zero, and, in particular for $\K^n=S^n$, its 
spectrum is given by%
\Eq{
k^2=l(l+n-1),\ l=0,1,2,\cdots.
}

The above definitions of $\SHB_i$ and $\SHB_{ij}$ become meaningless 
for $k^2=0$. Because the harmonic function for $k^2=0$ is  constant, 
the corresponding perturbation merely represents a change of the 
background metric with respect to a variation of the mass parameter. 
Thus, this is a trivial perturbation with respect to the uniqueness 
problem. Therefore, we only consider modes with $k^2>0$ from this 
point, unless otherwise stated.

In addition to $k^2=0$, harmonics with $k^2=n$ for $K=1$ are also 
exceptional, because $\SHB_{ij}$ vanishes for such harmonic 
functions. Modes corresponding to such harmonics are gauge modes as 
shown in Ref.~\citen{Kodama.H&Ishibashi2003}, and we do not consider 
such modes. This implies that $k^2>n$ can be assumed when $K=1$, 
because for $K=1$, the second smallest eigenvalue of $k^2$ is 
greater than or equal to 
$n$\cite{Gibbons.G&Hartnoll2002}.

Ignoring these exceptional modes, stationary scalar perturbations of 
the 
metric can be expanded in terms of the harmonics as
\Eq{
h_{ab}=f_{ab}(r)\SHB,\ 
h_{ai}=r f_a(r)\SHB_i,\ 
h_{ij}=2r^2\inpare{H_L(r)\gamma_{ij}\SHB+H_T(r)\SHB_{ij}}.
}
We adopt the following combinations as a basis for gauge-invariant 
quantities constructed from these expansion 
coefficients\cite{Kodama.H&Ishibashi2003}:
\Eqrsubl{GaugeInvariants:Scalar:static}{
&& F:= H_L + \frac{1}{n}H_T + \frac{f}{r}X_r,\\
&& F^t_t:=f^t_t + f' X_r,\\
&& F^r_r:=f^r_r + 2f X_r' + f' X_r,\\
&& F^r_t:=f^r_t+ f X_t' -f' X_t,
}
where
\Eq{
X_r:=\frac{r}{k}f_r+\frac{r^2}{k^2}H_T',\quad
X_t:=\frac{r}{k}f_t.
\label{Xa:static}
}
%

As shown in Ref. \citen{Kodama.H&Ishibashi2003}, the linearised 
Einstein equations for stationary scalar perturbations can be 
written in terms of these gauge-invariant variables as
\Eqrsubl{BasicEq:Scalar:static}{
&& F^r_r + F^t_t=-2(n-2)F,\\
&& F^r_t=0,\\
&& f'X'= \left(\frac{2(n-1)}{r^2}(f-K)+\frac{4(n+1)\lambda}{n}
         +\frac{2f'}{r}-\frac{(f')^2}{2f}+\frac{2f''}{n}\right)X \notag\\
&& \qquad -\left(\frac{2(f-K)}{r^2}
   +\frac{4(n^2-1)}{n}\lambda 
   +\frac{2(n-1)}{r}f'
   -\frac{(f')^2}{2f}
   \right. \notag\\
&& \quad \qquad \left. 
+\frac{2(n-1)}{n}f''-\frac{2(k^2-nK)}{r^2}\right)Y, 
\label{BasicEq:Scalar:static:1}\\
&& Y'= \frac{f'}{2f}(X-Y),
\label{BasicEq:Scalar:static:2}
}
where
\Eq{
X:=r^{n-2}(F^t_t-2F),\quad
Y:=r^{n-2}(F^r_r-2F).
}
%

As discussed in Ref. \citen{Kodama.H&Ishibashi2003}, we can reduce 
this set of equations to a second-order ODE in various ways. For 
example, the master variable $\Phi$ introduced in Ref. 
\citen{Kodama.H&Ishibashi2003} for generic perturbations can be 
written as a complicated linear combination of $X$ and $Y$ and 
satisfies a second-order ODE for stationary perturbations. However, 
this equation is not useful in the analysis of stationary 
perturbations, because its effective potential is not positive 
definite in the case of a non-vanishing cosmological constant for  
generic values of $n$. Further, it is not easy to determine the 
asymptotic behaviour of $F$ and $F^a_b$, because their expressions 
in terms of $\Phi$ are rather complicated. Therefore, in the present 
paper, we utilise second-order ODEs for $X$ and $Y$, which turn out 
to have structures convenient for the investigation of the 
uniqueness issue.  

The second-order ODE for $Y$ can be easily obtained by eliminating 
$X$ from \eqref{BasicEq:Scalar:static:1} with the help of 
\eqref{BasicEq:Scalar:static:2}\cite{Kodama.H&Ishibashi2003}. The 
result can be expressed as
\Eq{
\inpare{\frac{f^2}{r^{n-2}(f')^2} Y'}' =\frac{f U_Y}{r^n(f')^2}Y,
\label{MasterEq:Scalar:static:Y}
}
where
\Eq{
U_Y:=k^2-2(n-1)K+(n-2)f.
}
%
Similarly, by eliminating $Y$, we obtain the following second-order 
ODE for $X$: 
\Eq{
\inpare{\frac{f^2}{r^{n-4}P}X'}'=\frac{fU_X}{r^{n-2}P^2}X,
\label{MasterEq:Scalar:static:X}
}
where
\Eqrsub{
& P
 :=& 4\insbra{(n+1)x+m-K}f
   +\insbra{(n+1)x-2K}^2,\\
& U_X 
 :=& 4(n-2)\insbra{(n+1)(n+2)x+3(m-K)}f^2 \notag\\
&& +\Big[5(n-2)\inrbra{(n+1)x-2K}^2
     +4n(n+1)\inrbra{m-(n-2)K}x \notag\\
&& \qquad +4m\inrbra{m-(n-1)K}+4(n-2)K^2\Big] f \notag\\
&&+3\insbra{m-(n-2)K}\insbra{(n+1)x-2K}^2.
\label{U_X}
}
%

\section{Boundary Conditions}
\label{sec:BoundaryConditions}

In the arguments regarding black hole uniqueness, the boundary 
conditions at the horizon and at infinity play a crucial role. 
Obviously, if we impose boundary conditions that are too weak, the 
uniqueness will always be violated. On the other hand, the 
uniqueness theorem obtained under conditions that are too strong 
will not be sufficiently powerful in general. 

\subsection{Regularity condition at the horizon}

Because we are considering only regular black holes, the spacetime 
metric describing a black hole must be regular at the horizon. 
Hence, it is natural to require that a metric perturbation be 
regular at the horizon with respect to a coordinate system in which 
the background metric is regular. For the background metric 
\eqref{GeneralisedStaticMetric} with 
\eqref{GeneralisedStaticMetric:2Dpart}, such a coordinate system is 
given by the Szekeres-type coordinates $U,V$ and $z^i$ that satisfy
\Eq{
UV=-\frac{1}{\kappa^2}e^{2\kappa r_*},\quad
|V/U|=e^{2\kappa t},
}
where $\kappa=f'(r_h)/2$ is the surface gravity of the black hole, 
and $r_*$ is the coordinate defined by 
\Eq{
dr_*=\frac{dr}{f(r)}.
}
Because the black hole horizon is non-degenerate, we have 
$\kappa\not=0$, and $UV$ can be expressed near the horizon in terms 
of a regular positive function $g(r)$ as
\Eq{
UV=-f(r) g(r).
\label{UV:regularity}
}
In the present paper, we require that the components of 
perturbations of the metric and the Weyl tensors in this coordinate 
system be bounded at the horizon.

\subsubsection{Tensor perturbations}

For a tensor perturbation, the metric components in the 
$(U,V,z^i)$ coordinate system are bounded only when $H_T$ is bounded 
at the horizon. Under this condition, from \eqref{UV:regularity} and 
\eqref{deltaC:Tensor:UV}, we see that a perturbation of the Weyl 
tensor, $\delta C_{****}$, in this coordinate system is bounded if 
and only if $H_T'$ and $H_T''$ are bounded at $r=r_h$. Thus, the 
regularity condition at horizon is given by
\Eq{
H_T, H_T', H_T'' =\Order{1}.
\label{BCatHorizon:Tensor}
}
%

\subsubsection{Vector perturbations}

For a stationary vector perturbation, $F_t=f_t$ is the only 
non-vanishing gauge-invariant variable. Because we have 
\Eq{
f_U=\frac{f}{2\kappa U}(f^t+f_r),\quad
f_V=\frac{f}{2\kappa V}(-f^t+f_r),
\label{fainUV}
}
the metric components in the $(U,V,z^i)$ coordinate system are 
bounded only when $F^t=f^t$ is bounded at the horizon. Under this 
condition, from \eqref{deltaC:Vector:tr}, it follows that the Weyl 
tensor in this coordinate system is bounded at the horizon if and 
only if $F^{(1)}$ is bounded there. To summarise, the regularity 
condition is given  by
\Eq{
F^t, F^{(1)}=\Order{1}.
\label{BCatHorizon:Vector}
}
%

\subsubsection{Scalar perturbations}

For a scalar perturbation, the determination of the boundary 
condition at the horizon is not so simple, because all metric 
components can be non-vanishing. First, from the relations 
\Eqrsub{
&& f_{UU}=\frac{f^2}{4\kappa^2 U^2}(f^{tt}+f_{rr}-2f^t_r),\\
&& f_{VV}=\frac{f^2}{4\kappa^2 V^2}(f^{tt}+f_{rr}+2f^t_r),\\
&& f_{UV}=\frac{f}{4\kappa^2 UV} f^a_a,
}
we obtain the conditions
\Eq{
f^t_t, f^r_r, f^t_r=\Order{1},\quad
f^t_t-f^r_r=\Order{f}
}
at the horizon. From this, we have $X^t=\Order{1}$. Next, from 
\eqref{fainUV} and the regularity of $h_{ij}$, we obtain the 
conditions
\Eq{
f^t, f_r, H_T, H_L=\Order{1}
}
at the horizon. Under these conditions, the coefficients of 
$\SHB_{ij}$ in $\Lie_\eta C_{aibj}$ given in \eqref{LieC:Scalar} are 
bounded at the horizon. Therefore, the corresponding coefficients in 
$[\delta C_{aibj}]$ should be bounded in the 
$(U,V,z^i)$ coordinates. Taking account of \eqref{deltaC:Scalar:UV} 
and the equation 
$F^a_a=-2(n-2)F$, this leads to the conditions
\Eq{
F, F^a_a =\Order{1},\quad
F^t_t-F^r_r=\Order{f}
}
at the horizon. From this and the definitions of $F^a_b$, it follows that
\Eq{
X_r, X_r', H_T' =\Order{1}
}
at the horizon. Using \eqref{LieC:Scalar}, we can easily check  that 
all components of $\Lie_\eta C_{****}$ in the $(U,V,z^i)$ 
coordinates are bounded at the horizon under these boundary 
conditions. This implies that the gauge-invariant combinations 
$[\delta C_{****}]$ in the same coordinates should be bounded at the 
horizon. This requirement yields the additional conditions
\Eq{
F', (F^t_t)', (F^r_r)', f(F^t_t)'', F''=\Order{1}
}
at the horizon. 

In terms of the variables $X$ and $Y$, the boundary conditions at 
the horizon obtained by the above argument are expressed simply as
\Eq{
X, Y, X', Y'=\Order{1}.
\label{BCatHorizon:Scalar}
}
Note that the conditions on the second derivatives follow from these 
conditions, with the help of the Einstein equations 
\eqref{BasicEq:Scalar:static}. Note also that in the case  
$\Lambda>0$, the same conditions should be imposed at the 
cosmological horizon. 

\subsection{Asymptotic condition at infinity}

In the cases $\Lambda\le0$, the region outside the horizon extends 
to $r=\infty$. Therefore, we must impose some asymptotic condition. 
In the present paper, we require that all components of the metric 
perturbation $h_{\mu\nu}$ in the natural orthonormal basis of the 
background metric fall off at infinity, i.e., 
\Eqrsub{
&& f^t_t, f_{rt}, f^r_r \tend 0,\\
&& f^{-1/2}f_t, f^{1/2} f_r \tend 0,\\
&& H_T, H_L \tend 0.
}

First, for a tensor perturbation, this requirement gives the single condition
\Eq{
H_T \tend 0.
\label{AsymptoticCondition:Tensor}
}
Next, for a stationary vector perturbation, these conditions can be 
expressed in terms of the gauge-invariant variables $F_t=f_t$ and 
$F^{(1)}=-r(f_t/r)'$ as 
\Eq{
f^{-1/2}F_t , rf^{-1/2} F^{(1)} \tend 0.
\label{AsymptoticCondition:Vector}
}
Note that we do not need an asymptotic condition on $F_r$, because 
$F_r$ vanishes for a stationary perturbation.  

Finally, for a stationary scalar perturbation, the Einstein equation 
$F^a_a=-2(n-2)F$ yields 
\Eq{
\frac{2}{r^{n-2}}(r^{n-2}fX_r)'+f^a_a+2(n-2)\inpare{H_L+\frac{1}{n}H_T}=0.
}
From this, we obtain the asymptotic condition
\Eq{
\frac{f}{r}X_r \tend 0.
}
Hence, the above asymptotic conditions on the metric perturbations 
can be expressed in terms of the gauge-invariant variables $X$ and 
$Y$ as 
\Eq{
\frac{X}{r^{n-2}}, \frac{Y}{r^{n-2}} \tend 0.
\label{AsymptoticCondition:Scalar}
}
%

\subsection{Exceptional vector perturbations}
\label{subsec:ExceptionalMode2}

As shown in \S\ref{subsec:BasicEqs:Vector}, we have an exact general 
solution for the exceptional vector perturbation with $k_V^2=(n-1)K$ 
for $K\ge0$. There is a subtlety concerning the regularity of this 
solution. In the case $\Lambda\le0$, the above asymptotic condition 
is satisfied by \eqref{ExceptionalMode:ft} only for the choice 
$C=0$. The solution corresponding to this choice is obtained when we 
treat, for example, the Kerr metric with small angular momentum as a 
perturbed form of the Schwarzschild metric in the standard 
$(t,r,\theta,\phi)$ coordinates. However, under this gauge 
condition, the regularity condition at the horizon, 
\eqref{BCatHorizon:Vector}, is not satisfied for $J\not=0$. 
Nevertheless, this solution is regular at the horizon, because we 
can set $C=-J/(n+1)/r_h^{n+1}$ through a gauge transformation, for 
which the solution satisfies \eqref{BCatHorizon:Vector}. The 
apparent singular behaviour of the perturbation for the former gauge 
arises because the Killing vector $\partial_t$ vanishes at the 
bifurcation sphere of the horizons for the Schwarzschild metric, 
while it becomes space-like and does not vanish there for the Kerr 
metric. That is, the mapping used to compare the two metrics is 
singular at the bifurcation sphere. In contrast, for the latter 
gauge, which corresponds to the redefinition of the angular 
coordinate from $\phi$ to $\tilde\phi=\phi
 -\Omega_h t$, with $\Omega_h$ being the angular velocity of the 
horizon, $\partial_t$ becomes parallel to null generators of the 
horizon and vanishes at the bifurcation sphere. Hence, the mapping 
between two spacetimes is regular at the horizon. The situation in 
the $\Lambda>0$ case is similar. In this case, to prove the 
regularity of the solution, we have to employ different choices of 
$C$ at the black hole horizon and at the cosmological horizon.

In connection to the above consideration, we now give a comment on 
another type of exceptional mode. As is well known, the Kerr 
solution can be extended to a larger regular family by introducing 
the NUT parameter $\nu$. This family reduces to the Schwarzschild 
solution in the simultaneous limit $J\tend 0$ and $\nu\tend 0$. In 
particular, the Taub-NUT solution with $J=0$ expressed as
\Eq{
ds^2=-f(dt+2\nu\cos\theta d\phi)^2
 +(r^2+\nu^2)(d\theta^2+\sin^2\theta d\phi^2) +\frac{dr^2}{f}, 
\label{TaubNUTmetric}
}
with 
\Eq{
f=\frac{r^2-2Mr-\nu^2}{r^2+\nu^2},
}
can be treated as a perturbed form of the Schwarzschild solution 
when $|\nu|$ is small. This perturbation is of the vector type and 
given by $h_{ti}=-2\nu f \VHB_i$, with
\Eq{
\VHB_\theta=0,\quad \VHB_\phi=\cos\theta.
}
It is easy to see that this $\VHB_i$ satisfies the divergence-free 
condition and that $f_t=f/r$ is a solution to the Einstein equations 
satisfying the regularity condition at the horizon. (Of course, the 
NUT solution is not asymptotically flat and does not satisfy the 
above asymptotic condition at infinity.) Further, $\VHB_i$ is also 
harmonic. However, the corresponding eigenvalue is given by 
$k_V^2=-1$, i.e.,  $l=0$. This apparently contradicts our previous 
claim that $k_V^2>0$ for $K=1$. This contradiction arises simply 
because $\VHB^i$ is singular and not square integrable on $S^2$. 

This peculiarity of the solution arises from two pathological 
features of the Taub-NUT solution. Firstly, each $r=\const$ 
submanifold is diffeomorphic to $S^3$ for $\nu\not=0$. In 
particular, the causality condition is violated, and there exist 
closed time-like curves in the region satisfying$f(r)>0$,  
corresponding to the outside of a black hole.  Second, each 
$t=\const$ section of this $S^3$ is actually a cylinder with two 
boundaries for $f(r)>0$, although they shrink to points at the 
bifurcation surface of the horizons. Further, this section becomes 
time-like near the boundary. Due to these features, taking the limit 
$\nu\tend 0$ is a singular procedure. 

In the present paper, we only consider $L^2$-normalizable 
perturbations. Therefore, solutions with peculiarities like this 
Taub-NUT solution are excluded from the argument of perturbative 
uniqueness. 

\section{Asymptotically Flat Case}

The analysis of the perturbative uniqueness of asymptotically flat 
black hole solutions for $K=1$ and $\Lambda=0$ is rather 
straightforward, because the perturbation equations are exactly 
soluble. We mainly consider the case of a spherically symmetric 
black hole, i.e., $\K^n=S^n$, and only briefly mention generic 
Einstein cases. 

\subsection{Tensor perturbations}

In terms of the variable
\Eq{
x:=\pfrac{r_0}{r}^{n-1},
}
the master equation \eqref{MasterEq:Tensor:static} can be written 
\Eq{
\frac{d^2 H_T}{dx^2}+\frac{1}{x-1}\frac{dH_T}{dx}
+\frac{k_T^2+2}{(n-1)^2x^2(x-1)}H_T=0.
}
This equation is of the Fuchs type, and its general solution can be 
expressed in terms of hypergeometric functions as 
\Eq{
H_T=\frac{A}{r^{l+n-1}}F(\alpha,\alpha,2\alpha;x)
   +Br^l F(\alpha',\alpha',2\alpha';x),
}
where $l$ is a solution of the equation
\Eq{
k_T^2=l(l+n-1)-2,
}
and 
\Eq{
\alpha:=\frac{l+n-1}{n-1},\quad
\alpha':=-\frac{l}{n-1}.
}

For $\K^n=S^n$, $l$ takes the discrete values $l=2,3,\cdots$. Hence, 
if we require that $H_T$ be bounded at infinity, which is a weaker 
condition than \eqref{AsymptoticCondition:Tensor}, we have $B=0$. 
However, from the general formula 
\Eq{
F(\alpha,\beta,\gamma;1)
  =\frac{\Gamma(\gamma)\Gamma(\gamma-\alpha-\beta)}
    {\Gamma(\gamma-\alpha)\Gamma(\gamma-\beta)},
\label{Formula:HyperG(1)}
}
we find that
\Eq{
\lim_{x\tend 1-}F(\alpha,\alpha,2\alpha;x)=+\infty.
}
Therefore, there exists no stationary tensor perturbation of the 
Schwarzschild-Tangherlini solution that is bounded and regular 
outside the horizon.  

Contrastingly, for the case in which $\K^n$ is a generic Einstein 
space, the above general solution always falls off at infinity if 
the Lichnerowicz operator has an eigentensor for which $k_T^2<-2$. 
In this case, because the above equation for $H_T$ always has a 
solution that is regular at $x=1$, there exists a regular and 
bounded stationary tensor perturbation, and uniqueness does not 
hold. 

\subsection{Vector perturbations}

Because there exists no harmonic vector for which $k_V^2<n-1$, $l$ 
defined by
\Eq{
k_V^2=l(l+n-1)-1
}
can be assumed to satisfy $l\ge1$. Because the solutions for $l=1$, 
i.e. the exceptional modes, have been discussed in 
\S\S\ref{subsec:BasicEqs:Vector} and 
\ref{subsec:ExceptionalMode2}, we need only consider modes with 
$l>1$. 

For a generic mode, the Einstein equations 
\eqref{EinsteinEq:Vector:static} are equivalent to
\Eq{
\frac{dF^t}{dx^2}+\frac{2}{x-1}\frac{dF^t}{dx}
  -\frac{n(x -1)-(l-1)(l+n)}{(n-1)^2 x^2(x-1)}F^t=0,
}
where $x=(r_0/r)^{n-1}$, as previously. The general solution to this 
equation can be expressed in terms of hypergeometric functions as 
\Eq{
F^t
 = \frac{A}{r^{l+n-1}}F(a+2,b+1,a+b+2;x)
    +Br^l F(-a,1-b,-a-b;x),
}
where
\Eq{
a=\frac{l+1}{n-1},\quad b=\frac{l-1}{n-1}.
}
Note that when $a+b$ is a non-negative integer, $F(-a,1-b,-a-b;x)$ 
should be replaced by a function $\tilde F(x)$ that is regular at 
$x=0$ and $\tilde F(0)=1$.

If we require that $F^t$ be bounded at $r=\infty$, which is weaker 
than the asymptotic condition 
\eqref{AsymptoticCondition:Vector}, $B$ should vanish. Under this 
condition, from \eqref{Formula:HyperG(1)}, we find that $F^t$ 
diverges at the horizon. Therefore, there exists no regular and 
bounded stationary vector perturbation other than the exceptional 
modes.

\subsection{Scalar perturbations}

As explained in \S\ref{subsec:BasicEqs:Scalar}, we can assume that 
$m=k^2-n>0$ for $K=1$. Therefore, we can express $k^2$ in terms of 
$l>1$ as 
\Eq{
k^2=l(l+n-1).
}
Then, \eqref{MasterEq:Scalar:static:Y} can be written in terms of $x=(r_0/r)^{n-1}$ as
\Eq{
\frac{d^2Y}{dx^2}+p\frac{dY}{dx}+q Y=0,
}
where
\Eq{
p=\frac{2(n-2)x+2}{(n-1)x(x-1)},\quad
q=\frac{-(l-1)(l+n)+(n-2)x}{(n-1)^2x^2(1-x)}.
}
This equation can be solved explicitly again, and its general solution is given by
\Eq{
Y=\frac{A r_0^{n-1}}{r^{n+l}}F_1(x) +B r_0^{n-1} r^{l-1} F_2(x),
}
with
\Eqrsub{
&& F_1(x)=F(\nu+1,\nu+2,2\nu+2;x),\\
&& F_2(x)=F(-\nu,-\nu+1,-2\nu;x),
}
where $\nu=l/(n-1)$, and $A$ and $B$ are arbitrary constants. 
Inserting this solution for $Y$ into 
\eqref{BasicEq:Scalar:static:2}, we obtain 
\Eqr{
& X=
 & -\frac{A}{n-1}\frac{1}{r^{l+1}}\left[\{2+(n-3)x\}F_1(x)
     +2(n+l-1)F_3(x)\right]\notag\\
&&-\frac{B}{n-1}r^{n+l-2}\left[\{2+(n-3)x\}F_2(x)
     -2lF_4(x)\right],
}
with
\Eqr{
&& F_3(x)=F(\nu,\nu+2,2\nu+2;x),\\
&& F_4(x)=F(-\nu-1,-\nu+1,-2\nu;x).
}

From these expressions, we find that the asymptotic condition 
\eqref{AsymptoticCondition:Scalar} is satisfied only when $B=0$ if 
$l>1$. However, when $B=0$, $Y$ diverges at $x=1$, and the 
regularity condition \eqref{BCatHorizon:Scalar} is not satisfied. 
Actually, the same conclusion is obtained from the weaker condition 
that $F$ and $F^a_b$ are bounded at infinity. Hence, there exists no 
regular and bounded stationary scalar perturbation.   

The following proposition summarises the main result obtained in 
this section. 

\medskip
\begin{proposition}\em
For the Schwarzschild-Tangherlini black hole, there exists no 
stationary perturbation that is regular and bounded outside the horizon, 
except for the exceptional vector perturbations.
\end{proposition}
\medskip

Taking account of the uniqueness theorem for asymptotically flat 
static vacuum regular solutions in higher 
dimensions\cite{Hwang.S1998,Gibbons.G&Ida&Shiromizu2002a}, this 
result implies that the Myers-Perry solution is the only regular 
stationary solution for an asymptotically flat vacuum system near 
the static limit in a spacetime with any number of dimensions. 
However, for a generic system for which $\K^n$ is a non-spherically 
symmetric Einstein space with $K=1$, there may exist a regular 
static tensor perturbation that falls off at infinity.  

\section{Asymptotically de Sitter Case}

In the case with a  non-vanishing cosmological constant $\Lambda$, 
we cannot express general solutions to the perturbation equations in 
terms of known functions explicitly. Nevertheless, we can show the 
non-existence of stationary perturbations satisfying the boundary 
conditions with the help of integral identities derived from the 
master equations. 

We first consider the case $\Lambda>0$, for which $K$ must be unity. 
In this case, we can demonstrate uniqueness only by studying the 
behaviour of perturbations in the region bounded by the black hole  
horizon $r=r_h$ and the cosmological horizon $r=r_c$, for which only 
the regularity condition at the horizon is relevant. 

\subsection{Tensor perturbations}

For a tensor perturbation, by multiplying 
\eqref{MasterEq:Tensor:static} by $r^{n/2}H_T$, integrating it over 
$r$, and using the regularity condition 
\eqref{BCatHorizon:Tensor}, we obtain the identity
\Eq{
0=\insbra{fr^{n/2}H_T\inpare{r^{n/2}H_T}'}_{r_h}^{r_c}
=\int_{r_h}^{r_c} dr\insbra{f\{(r^{n/2}H_T)'\}^2
   +r^{n-2} U_T H_T^2}.
}
When $\K^n=S^n$, $U_T$ is always positive 
in the range $r_h<r<r_c$, because $k_T^2\ge n-2$. Hence, from this 
integral identity, it follows that there exists no regular 
stationary tensor perturbation.

By contrast, in the case in which $\K^n$ is not a constant curvature 
space, $k_T^2$ may become negative, and no general lower bound on it 
is known. Therefore, there may exist a regular stationary 
perturbation.

\subsection{Vector perturbations}

The argument concerning vector perturbations is almost the same as 
that for tensor perturbations. We multiply 
\eqref{EinsteinEq:Vector:static:1} by $F_t/r$ and integrate it over 
$r$. Then, using the regularity condition 
\eqref{BCatHorizon:Vector}, we obtain
\Eq{
0=-\insbra{r^n f F^t F^{(1)}}_{r_h}^{r_c}
 = \int_{r_h}^{r_c}dr\, \insbra{r^n\inpare{F^{(1)}}^2
  +m_V r^{n-2}f \inpare{F^t}^2 }.
}
As in the case $\Lambda=0$, we can restrict our consideration to the 
case in which $m_V=k_V^2-n+1>0$. Then, only the trivial solution 
$F^t=0$ satisfies this integral identity. Unlike the case of tensor 
perturbations, this result holds even if $\K^n$ is a generic 
Einstein space with $K=1$.

\subsection{Scalar perturbations}

The argument for a scalar perturbation is also similar to that given 
above. Now, we utilise the master equation for $X$, 
\eqref{MasterEq:Scalar:static:X}, which leads to the identity
\Eq{
\insbra{\frac{f^2}{r^{n-4}P}XX'}_{r_h}^{r_c}
 =\int_{r_h}^{r_c}dr\insbra{\frac{f^2}{r^{n-4}P}\inpare{X'}^2
   +\frac{fU_X}{r^{n-2}P^2}X^2}.
}
For $f=1-x-\lambda r^2>0$, $P$ vanishes only when $m+1=0$, which is 
not realised, because $m=k^2-n>0$. Further, if $x=2/(n+1)$ at 
$r=r_h$ or $r=r_c$, $P'$ does not vanish there. Hence, $f^2/P$ is 
positive for $r_h<r<r_c$ and vanishes at $r=r_h$ and $r_c$. From 
this and the regularity condition \eqref{BCatHorizon:Scalar}, we 
find that the left-hand side of the above integral identity 
vanishes. Therefore, if $U_X$ is non-negative, we can conclude that 
there exists no stationary regular scalar perturbation. From 
\eqref{U_X}, we see that this condition is satisfied if $m\ge n-1$. 
Although this inequality may not hold for a generic Einstein space 
$\K^n$ with $K=1$, in the most important case, in which $\K^n=S^n$, we have the necessary inequality, because $m$ takes the discrete values 
$m=(l-1)(l+n)$ with $l\ge2$:
\Eq{
m-(n-1)=(l-2)(l+n+1)+3>0.
}

To summarise, we have proved the following proposition.
\medskip
\begin{proposition}\em
For a de Sitter-Schwarzschild black hole, there exists no stationary 
perturbation that is regular in the region $r_h\le r\le r_c$, except 
for the exceptional vector perturbations.
\end{proposition}
\medskip

From the above arguments, it follows that near the static and 
spherically symmetric limit, the rotating version of the de 
Sitter-Schwarzschild solution with the same number of parameters as 
the Myers-Perry solution is the only regular stationary vacuum 
solution for $\Lambda>0$. In four dimensions, this solution is 
identical to the Carter solution with a vanishing NUT parameter, and 
in higher dimensions, it is identical to the Gibbons-L\"u-Page-Pope 
solution with $\Lambda>0$. This uniqueness may not hold when 
$\K^n$ is not a constant curvature space, due to the existence of a 
regular static perturbation of the tensor or scalar type in that 
case.

\section{Asymptotically anti-de Sitter Case}

The method based on integral identities can also be applied to the 
case $\Lambda<0$. However, because the region outside the horizon 
extends to infinity in this case, we have to check that the boundary 
contribution at infinity vanishes for solutions satisfying the 
asymptotic condition. Further, we also have to consider the cases  
$K=0$ and $K=-1$.

\subsection{Tensor perturbations}

For large $r$, the master equation \eqref{MasterEq:Tensor:static} 
can be approximated by
\Eq{
H_T'' +\frac{n+2}{r}H_T' \approx 0,
}
whose general solution can be written  $A/r^{n+1}+B$, with constants 
$A$ and $B$. Hence, if we require the asymptotic condition 
\eqref{AsymptoticCondition:Tensor}, $H_T$ should behave as 
$\sim 1/r^{n+1}$ at infinity. From this and the regularity condition 
at the horizon, \eqref{BCatHorizon:Tensor}, it follows that the 
left-hand side of the identity
\Eq{
\insbra{fr^{n/2}H_T\inpare{r^{n/2}H_T}'}^{\infty}_{r_h}
=\int_{r_h}^{\infty} dr\insbra{f\{(r^{n/2}H_T)'\}^2
   +r^{n-2} U_T H_T^2}
}
vanishes. Hence, in the case in which $\K^n$ is a constant curvature 
space, from the positivity of $U_T$, we can conclude that there 
exists no regular stationary tensor perturbation that satisfies the 
fall off condition \eqref{AsymptoticCondition:Tensor}.

\subsection{Vector perturbations}

For large $r$, the master equation \eqref{EinsteinEq:Vector:static} 
can be approximated by
\Eq{
\insbra{r^{n+2}\inpare{r F^t}'}'\approx0,
}
whose general solution can be written $A/r^{n+2}+B/r$. Because 
$f^{-1/2}F_t$ behaves as $\sim r F^t$ at infinity in an 
asymptotically anti-de Sitter spacetime, the asymptotic condition 
\eqref{AsymptoticCondition:Vector} requires $B$ to vanish. Then, in 
the identity obtained from \eqref{EinsteinEq:Vector:static}, 
\Eq{
-\insbra{r^n f F^t F^{(1)}}_{r_h}^{\infty}
 = \int_{r_h}^{\infty}dr\, \insbra{r^n\inpare{F^{(1)}}^2
  +m_V r^{n-2}f \inpare{F^t}^2 },
}
the boundary terms on the left-hand side vanish if the regularity 
condition \eqref{BCatHorizon:Vector} at the horizon is also 
satisfied. From this, it follows that $F^t$ should vanish 
identically if $m_V>0$. Therefore, the exceptional modes discussed 
in 
\S\ref{subsec:ExceptionalMode2} are the only regular stationary 
vector perturbations that satisfy the fall-off condition 
\eqref{AsymptoticCondition:Vector}, irrespective of the value of 
$K$. 

\subsection{Scalar perturbations}

In order to determine the asymptotic behaviour of scalar 
perturbations for $\Lambda<0$, we utilise the equation for $Y$, 
\eqref{MasterEq:Scalar:static:Y}. For large $r$, this equation can 
be approximated as 
\Eq{
Y'' -\frac{n-4}{r}Y' -\frac{n-2}{r^2}Y \approx0.
}
The general solution to this equation can be expressed as
\Eq{
Y \approx \frac{A}{r} + B r^{n-2},
}
and the asymptotic condition \eqref{AsymptoticCondition:Scalar} is 
satisfied only if $B=0$. For this solution, the boundary terms in the 
identity obtained from \eqref{MasterEq:Scalar:static:Y},
\Eq{
\insbra{\frac{f^2}{r^{n-2}(f')^2}Y Y'}_{r_h}^{\infty}
=\int_{r_h}^\infty dr \frac{1}{r^{n}(f')^2}
 \insbra{r^2f^2 (Y')^2+ fU_Y Y^2},
}
vanish if the solution also satisfies the regularity condition 
\eqref{BCatHorizon:Scalar} at the horizon. Thus, uniqueness holds if 
$U_Y$ is non-negative. For $K\le0$, this condition is trivially 
satisfied. In contrast, for $K=1$, it leads to the condition on the 
spectrum $k^2-2(n-1)\ge0$. As in the case $\Lambda\ge0$, this 
condition is satisfied for $\K^n=S^n$, but it may not be satisfied 
generically.

The main result obtained in this section can be summarised as the following proposition.

\medskip
\begin{proposition}\em
For anti-de Sitter-Schwarzschild black holes and topological black 
holes, there exists no stationary perturbation that is regular and 
falls off at infinity, except for the 
exceptional vector perturbations. 
\end{proposition}
\medskip

This result indicates that the Gibbons-L\"u-Page-Pope solution with 
$\Lambda<0$, which is a rotational extension of the anti-de 
Sitter-Schwarzschild solution characterised by the same number of 
parameters as for the Myers-Perry solution, is the only regular 
stationary asymptotically anti-de Sitter solution near the static 
and spherically symmetric limit. Concerning topological black holes 
with $K=-1$, no such rotational 
extension exists if $\K^n$ is compact, because there is no 
exceptional mode in this case (cf. Refs. \citen{Klemm.D&Moretti&Vanzo1998} and \citen{Lemos.J1995}). This uniqueness may be violated for 
the case in which $\K^n$ is a generic Einstein space with a 
non-constant sectional curvature, due to the existence of a regular 
static perturbation of tensor type for $K=0,\pm1$ or of scalar type 
for $K=1$.

\section{Concluding Remarks}

In the present paper, we have determined all stationary solutions to 
the perturbation equations that are regular at the horizon and fall 
off at infinity (in the case $\Lambda\le0$) for a static black hole 
background whose horizon has a spatial section with constant 
curvature. As summarised in Theorem 1 given in the introduction, we 
have found that these solutions are exhausted by the trivial 
perturbations corresponding to variations of the background 
parameters and the exceptional vector perturbations representing 
stationary rotations of black holes for any value of the 
cosmological constant. We have also pointed out that there may exist 
additional regular and bounded solutions if a spatial section of the 
horizon does not have constant curvature. 

As mentioned in the introduction, this result strongly indicates 
that the Myers-Perry solution and the Gibbons-L\"u-Page-Pope 
solution are the only asymptotically flat, de Sitter or anti-de 
Sitter regular stationary vacuum solutions near the static and 
spherically symmetric limit. However, our arguments do not give an 
exact proof of this perturbative uniqueness near the static and 
spherically symmetric limit, because there may exist a regular 
family of solutions that approaches the limit in a singular way, as 
in the case of the Taub-NUT solution discussed in 
\S\ref{subsec:ExceptionalMode2}. Logically, there is also the 
possibility that there exist infinitely many families whose 
bifurcation points accumulate at the static limit, although this is 
quite unlikely. 

It is clear that we have to study perturbations of the Myers-Perry 
solutions in order to show that such a pathological situation is not 
realised and to determine whether the Myers-Perry family actually 
bifurcates at high angular momenta, as suggested by Emparan and 
Myers\cite{Emparan.R&Myers2003}. At present, however, such a study 
would be quite difficult because there exists no tractable 
formulation for perturbations of the Myers-Perry solutions. The 
development of such a formulation is the most important problem to 
be solved in the present context.

Finally, we would like to point out that the fall-off condition at 
infinity is essential for uniqueness to hold in the asymptotically 
anti-de Sitter case, unlike in the asymptotically flat case, in 
which uniqueness holds even if we only require perturbations to be 
bounded at infinity. In fact, for the general solution given in the 
previous section, $\delta g_{\mu\nu}$ with respect to the natural 
orthonormal frame is bounded at infinity even if $B\not=0$, 
irrespective of the type of perturbations. Further, we can easily 
show that one solution is always regular at the horizon for each 
mode. Hence, in the asymptotically anti-de Sitter case, for any 
boundary value of $\delta g_{\mu\nu}$ at infinity, there exists a 
stationary solution to the perturbation equation that is regular 
everywhere and satisfies the given boundary condition at infinity. 
Furthermore, this solution is unique except for the freedom to add 
solutions corresponding to the exceptional vector perturbations. 
This result is 
consistent with the general results concerning asymptotically 
anti-de Sitter and static solutions given in Refs. 
\citen{Anderson.M&Chrusciel&Delay2002} and 
\citen{Anderson.M&Chrusciel&Delay2004A}, and has a close connection 
with the AdS/CFT argument\cite{Maldacena.J1998,Anderson.M2004A}. 

\section*{Acknowledgements}
The author would like to thank Akihiro Ishibashi for valuable 
comments and Yoshiyuki Morisawa and Daisuke Ida for useful 
conversations. This work is supported by the JSPS grant No. 15540267.

\appendix
\section{Interpretation of exceptional modes for vector 
perturbations}\label{Appendix:ExceptionalModes}

In this appendix, we derive the relation between the angular 
momentum of the system and the parameter $J$ for each exceptional 
mode of the vector perturbation in \S\ref{subsec:BasicEqs:Vector}.

To begin, we note that for an exceptional mode, the metric can be 
written  
\Eq{
ds^2=ds_0^2 + 2r f_t(r) \VHB_i dt dz^i,
}
where $ds_0^2$ is the background metric. Because $\VHB_i$ is a 
Killing vector for $k_V^2=(n-1)K$\cite{Kodama.H&Ishibashi2004}, this 
perturbed metric is  invariant under translations generated by the 
vector field 
\Eq{
\eta:= \VHB^i \partial_i.
}
Hence, it is ``rotationally symmetric.'' This implies that we can 
calculate the angular momentum of this spacetime using the Komar 
integral on $\K^n$. In fact, from the Einstein equations for the 
background metric 
\Eq{
R_{\mu\nu}=(n+1)\lambda g_{\mu\nu},
}
we obtain the identity
\Eq{
d * d\eta_*= 2(n+1)\lambda I_\eta \Omega_{n+2}
}
for any Killing vector $\eta$, where $\eta_*=\eta_\mu dx^\mu$, $*$ 
is the Hodge dual operator, $I_\eta$ is the inner product operator, 
and $\Omega_{n+2}$ is the spacetime volume form. From this, it 
follows that the integration of the $n$-form $*d\eta_*$ over a 
$n$-subspace $\Sigma_n$ is independent of the choice of $\Sigma_n$ 
if $\Sigma_n$ is tangential to the vector field $\eta$. If we 
calculate this integral in the present case in which
\Eq{
\eta_*=\frac{J}{n+1}\frac{\VHB\cdot\VHB}{r^{n-1}} dt,
}
we obtain
\Eq{
\int_{\Sigma_n} *d\eta_*= C_{\VHB} \frac{n-1}{n+1} J,
\label{KomarIngtegral}
}
where
\Eq{
C_{\VHB}:= \int_{\K^n} \VHB\cdot \VHB d\Omega_n;\quad
d\Omega_n= \sqrt{\gamma}d^n z.
}
Therefore, $J$ is proportional to the angular momentum of the 
spacetime defined by the Komar integral.

In order to determine the proportionality constant in this relation, 
we utilise the perturbation equation with a source term for an 
exceptional mode derived in Ref. \citen{Kodama.H&Ishibashi2004}. In 
the stationary case, it reads
\Eq{
\inpare{r^{n+1}F^{(1)}}'=2\kappa^2 r^{n+1}\tau^t,
\label{VectorEqWithSource}}
where $\tau^t$ is related to the energy-momentum tensor of the 
source as
\Eq{
T^t_i=r\tau^t \VHB_i.
}
In a flat background, the angular momentum $\bm{J}[\VHB]$ of the 
system with respect to the ``rotational'' Killing vector $\VHB^i$ 
can be expressed as
\Eq{
\bm{J}[\VHB]=\int_0^\infty dr r^n \int_{\K^n}d\Omega_n T_{ti}\VHB^i.
}
Therefore, integration of \eqref{VectorEqWithSource} over $r$ yields
\Eq{
(r^{n+1}F^{(1)})(r=\infty)=-\frac{2\kappa^2}{C_\VHB}
   \bm{J}[\VHB].
}
Comparing this expression and \eqref{ExpectionalMode:F1}, we obtain
\Eq{
J=\frac{2\kappa^2}{C_\VHB} \bm{J}[\VHB].
}
This final result can be regarded as exact, because it is consistent 
with the expression in terms of the Komar integral, 
\eqref{KomarIngtegral}, and it gives the relation
\Eq{
\bm{J}[\VHB]=\frac{1}{2\kappa^2}\frac{n+1}{n-1}\int_{\Sigma_n} 
*d\eta_*.
}
%

\section{$\delta C_{****}$}

In this appendix, we derive explicit expressions for the background 
values and the perturbation of the Weyl curvature.

\subsection{$C_{****}$}
The Weyl tensor for our static background metric is given by
\Eqrsub{
&& C_{abcd}=\frac{n(n-1)}{2(n+1)}\Psi (g_{ac}g_{bd}-g_{ad}g_{bc}),\\
&& C_{aibj}=-\frac{n-1}{2(n+1)}r^2 \Psi g_{ab}\gamma_{ij},\\
&& C_{ijkl}=\frac{r^4\Psi}{n+1}(\gamma_{ik}\gamma_{jl}
                -\gamma_{il}\gamma_{jk}),
}
where
\Eq{
\Psi:= \frac{\Box r}{r}+ 2\frac{K-(Dr)^2}{r^2}
 =\frac{(n+1)r_0^{n-1}}{r^{n+1}}.
}
%

\subsection{Tensor perturbations}

For a tensor perturbation, all components of the perturbation of the 
Weyl tensor, $\delta C_{****}$, are gauge invariant. For a solution 
to \eqref{MasterEq:Tensor:static}, their non-vanishing 
components in the $(t,r,z^i)$ coordinate system are given by
\Eqrsubl{deltaC:Tensor:tr}{
& \frac{1}{r^2 f}\delta C_{t i t j}=
  & \left[\frac{f'}{2} H_T'
    +\frac{n-1}{n+1}\Psi H_T\right] \THB_{ij},
\\
& \frac{f}{r^2}\delta C_{r i r j}=
  & \Big[\left(-(n-1)\lambda +\frac{(n-2)K}{r^2}
    -\frac{n-3}{2(n+1)}\Psi
    \right)r H_T' 
\notag\\
&&\quad  -\left(\frac{k^2+2K}{r^2} 
    +\frac{n-1}{n+1}\Psi\right)H_T\Big] \THB{ij},
\\
& \frac{\sqrt{f}}{r^3}\delta C_{r ijk}=
  & \frac{\sqrt{f}}{r}  H_T' 
    \left(\hat D_k \THB_{ij}-\hat D_j \THB_{ik}\right),
\\
& \frac{1}{r^4}\delta C_{ijkl}=
  & \left[-\frac{f}{r} H_T'
         +\left(\frac{2\Psi}{n+1}-\frac{K}{r^2}\right)H_T
    \right] \THB^{(0)}_{ijkl}
    +\frac{H_T}{r^2} \THB^{(1)}_{ijkl}.
}
%
From these, we find that the non-vanishing components of $\delta 
C_{****}$ in the $(U,V,z^i)$ coordinate system are given by
\Eqrsubl{deltaC:Tensor:UV}{
& \delta C_{UiUj}
 =&\frac{f^2}{4\kappa^2 U^2}\insbra{
    -r^2 H_T''  -2 r H_T'}\THB_{ij},\\
& \delta C_{ViVj}
 =&\frac{f^2}{4\kappa^2 V^2}\insbra{
    -r^2 H_T''  -2r H_T'}\THB_{ij},\\
& \delta C_{UiVj}
 =& \frac{r^2f}{4\kappa^2 UV}
 \insbra{\inpare{f'+\frac{n-2}{r}f} H_T'
    +2\frac{n-1}{n+1}\Psi H_T }\THB_{ij},\\
& \delta C_{Uijk}
 =&\frac{r^2f}{2\kappa U} H_T'(\hat D_k \THB_{ij}-\hat 
D_j\THB_{ik}),\\
& \delta C_{Vijk}
 =&\frac{r^2f}{2\kappa V} H_T'(\hat D_k \THB_{ij}-\hat 
D_j\THB_{ik}).
}
The expression for $\delta C_{ijkl}$ is the same as that above.
 
\subsection{Vector perturbations}

For a vector perturbation, $\delta C_{****}$ are not gauge 
invariant. Under the vector-type coordinate transformation
\Eq{
\xi^a:=\bar\delta y^a=0,\quad \xi^i:=\bar\delta z^i=r^2 L\VHB^i,
\label{GaugeTrf:Vector}
}
they transform as
\Eq{
\bar\delta (\delta C_{\mu\nu\lambda\sigma})=-\Lie_\xi 
C_{\mu\nu\lambda\sigma}.
}
Hence, taking account of the fact that $H_T$ transforms under 
\eqref{GaugeTrf:Vector} as
\Eq{
\bar \delta H_T=kL,
}
we find that the following combination is gauge invariant:
\Eq{
[\delta C_{\mu\nu\lambda\sigma}]
:=\delta C_{\mu\nu\lambda\sigma}+\Lie_\eta C_{\mu\nu\lambda\sigma},
\label{deltaC:gauge-invariant:def}
}
where
\Eq{
\eta_a=0,\quad \eta_i=\frac{r^2}{k}H_T \VHB_i.
}
%

Explicitly, the non-vanishing components of $\Lie_\eta 
C_{\mu\nu\lambda\sigma}$ are as follows:
\Eqrsubl{LieC:Vector}{
&& \Lie_\eta C_{abci}
 = -\frac{n-1}{2(n+1)}\frac{r^2}{k}\Psi
       (g_{ac}D_b H_T-g_{bc}D_a H_T)\VHB_i,\\
&& \Lie_\eta C_{aibj}
 = \frac{n-1}{n+1}r^2 \Psi H_T \VHB_{ij},\\
&& \Lie_\eta C_{aijk}
 =-\frac{r^4}{(n+1)k}\Psi D_a H_T
    (\gamma_{ij}\VHB_k-\gamma_{ik}\VHB_j), \\
&& \Lie_\eta C_{ijkl}
 = -\frac{2r^4}{n+1}\Psi H_T 
  (\gamma_{ik}\VHB_{jl}+\gamma_{jl}\VHB_{ik}
          -\gamma_{il}\VHB_{jk}-\gamma_{jk}\VHB_{il}).
}
%

In the $(t,r,z^i)$ coordinate system, the non-vanishing components 
of $[\delta C_{****}]$ are given by
\Eqrsubl{deltaC:Vector:tr}{
& [\delta C_{rtri}]
  =& 
  \Big[-(n-1)F^{(1)}
  +\frac{F^t}{r}\Big\{\frac{n^2+3n-2}{4n(n+1)}r^2\Psi\notag\\
&&\quad 
  +\frac{n^2-5n+2}{2n(n+1)}K-\frac{n-1}{n}m_V 
  +\frac{n^2+7n-2}{2n(n+1)}f
  \Big\}\Big]\VHB_i,\\
& [\delta C_{rtij}]
 =&-\frac{r}{2} F^{(1)} d\VHB_{ij},\\
& [\delta C_{tirj}]
 =&-\frac{r}{4} F^{(1)} d\VHB_{ij},\\
& [\delta C_{tijk}]
 =&-\frac{rf}{2}F^t \hat D_i d\VHB_{jk}
  +\frac{rf}{2n}\Big[-\frac{-(2n-1)r^2\Psi+2(n^2+n-1)K+2f}{n+1}F^t
  \notag\\
&&\quad +rF^{(1)}\Big](\gamma_{ij}\VHB_k-\gamma_{ik}\VHB_{j}),
}
where 
\Eq{
d\VHB_{ij}:=\hat D_i \VHB_j -\hat D_j \VHB_i.
}
Note that from \eqref{LieC:Vector}, we have $[\delta 
C_{\mu\nu\lambda\sigma}]=\delta C_{\mu\nu\lambda\sigma}$ for these 
components.
The components in the $(U,V,z^i)$ coordinate system are expressed in 
terms of these non-vanishing components in the $(t,r,z^i)$ 
coordinate system as
\Eqrsub{
&& \frac{1}{V}[\delta C_{UVUi}]
  =\frac{1}{U}[\delta C_{UVVi}]
  =\frac{1}{4\kappa^3}\pfrac{f}{UV}^2[\delta C_{rtri}],\\
&& [\delta C_{UVij}]=2[\delta C_{UiVj}]
  =\frac{1}{2\kappa^2}\frac{f}{UV}[\delta C_{rtij}],\\
&& \frac{1}{U}[\delta C_{Vijk}]
  =-\frac{1}{V}[\delta C_{Uijk}]
  =\frac{1}{2\kappa}\frac{1}{UV}[\delta C_{tijk}].
}
%

\subsection{Scalar perturbations}

For a scalar perturbation, under the gauge transformation
\Eq{
\xi_a=T_a\SHB,\quad
\xi_i=r^2L \SHB_i,
}
$X_a$ and $H_T$ transform as
\Eq{
\bar\delta X_a=T_a,\quad
\bar\delta H_T =kL.
}
Hence, the combinations \eqref{deltaC:gauge-invariant:def} with
\Eq{
\eta_a=X_a \SHB,\quad
\eta_i=\frac{r^2}{k}H_T\SHB_i
}
are gauge invariant. 

Explicitly, the non-vanishing components of $\Lie_\eta 
C_{\mu\nu\lambda\sigma}$ are given by
\Eqrsubl{LieC:Scalar}{
&& \Lie_\eta C_{trtr}=
   \frac{n(n-1)}{2(n+1)}\Psi \insbra{-2D\cdot X
     +\frac{n+1}{r}Dr \cdot X} \SHB,
\\
&& \Lie_\eta C_{trci}=
   \frac{n-1}{2(n+1)k}\Psi \epsilon_{ca}
   \inpare{nk^2 X^a + r^2 D^a H_T} \SHB_i,
\\
&& \Lie_\eta C_{aibj}=
   \frac{(n-1)r \Psi}{2(n+1)}\left[(n-1)g_{ab}Dr\cdot X
        -r(D_a X_b+D_b X_a)\right]\gamma_{ij}\SHB
\notag\\
&&\qquad
  -\frac{n-1}{n(n+1)}kr^2\Psi H_T g_{ab}(n\SHB_{ij}+\gamma_{ij}\SHB),
\\
&& \Lie_\eta C_{aijk}=
   -\frac{r^2 \Psi}{2(n+1)k}\left[k^2(n-1)X_a +2r^2 D_a H_T\right]
        (\gamma_{ij}\SHB_k-\gamma_{ik}\SHB_j),
\\
&& \Lie_\eta C_{ijkl}=
   \frac{n-3}{n+1}r^3 \Psi Dr\cdot X (\gamma_{ik}\gamma_{jl}
                -\gamma_{il}\gamma_{jk})\SHB 
\notag\\
&&\qquad +\frac{2r^4 k}{n(n+1)}\Psi H_T \left[
         2n(\gamma_{i[k}\SHB_{l]j}-\gamma_{j[k}\SHB_{l]i})
         +2(\gamma_{ik}\gamma_{jl}-\gamma_{il}\gamma_{jk})\SHB 
         \right] .
}
%

The non-vanishing components of the gauge-invariant combinations in 
the $(t,r,z^i)$ coordinates are
\Eqrsubl{deltaC:Scalar:tr}{
& [\delta C_{rtrt}]
 =&\insbra{
  -\inpare{\frac{n^2-n+2}{4(n+1)}\Psi+\frac{f-K}{2r^2}}F^c_c 
 +2(fF^t_t)''-(f'F^t_t)'}\SHB,\\
& [\delta C_{rtti}]
 =& \frac{kf}{2}\insbra{(F^t_t)'-\frac{1}{r}F^t_t
      +\frac{f'}{2f}(F^t_t-F^r_r)}\SHB_i,\\
& [\delta C_{titj}]
 =& f\Big[\inrbra{\frac{n-1}{n+1}r^2\Psi+(n-2)rf'}F
    +\frac{1}{2}r^2f'F'-\frac{r}{2}F_{tt}' \notag\\
&& \quad -\inpare{\frac{k^2}{2n}-\frac{n-1}{2(n+1)}r^2\Psi}F^t_t
   \Big] \gamma_{ij}\SHB
   +\frac{k^2}{2}f F^t_t \SHB_{ij},\\
& [\delta C_{rirj}]
 =& \frac{1}{f}\Big[-r^2f F'' -\inpare{\frac{r^2}{2}f'+2rf}F'
    -\frac{n-1}{n+1}r^2\Psi F +\frac{r}{2}(fF^r_r)'\notag\\
&& \quad +\inpare{\frac{k^2}{2n}-\frac{n-1}{2(n+1)}r^2\Psi}F^r_r
   \Big] \gamma_{ij}\SHB
   -\frac{k^2}{2f} F^r_r \SHB_{ij},\\
& [\delta C_{rijk}]
 =& \frac{kr}{2}[2rF'-F^r_r] (\gamma_{ik}\SHB_j-\gamma_{ij}\SHB_k),\\
& [\delta C_{ijkl}]
 =& \frac{r^2}{n^2(n+1)}\Big[-2n^2(n+1)rfF'
    +2n\inrbra{(n+1)m+2nr^2\Psi}F \notag\\
&& \quad +n^2(n+1)f F^r_r\Big](\gamma_{ik}\gamma_{jl}
    -\gamma_{il}\gamma_{jk})\SHB \notag\\
&& -\frac{k^2r^2}{2}F \inpare{\gamma_{ik}\SHB_{jl}
   +\gamma_{jl}\SHB_{ik}-\gamma_{il}\SHB_{jk}-\gamma_{jk}\SHB_{il}}.
}
The corresponding components in the $(U,V,z^i)$ coordinates are
\Eqrsubl{deltaC:Scalar:UV}{
&& [\delta C_{UVUV}]
 =\frac{1}{4\kappa^4}\pfrac{f}{UV}^2 [\delta C_{rtrt}],\\
&&\frac{f}{U} [\delta C_{UVVi}]=-\frac{f}{V}[\delta 
C_{UVUi}]=\frac{1}{4\kappa^3}\pfrac{f}{UV}^2 [\delta C_{rtti}],
\\
&& \frac{4\kappa^2 U^2}{f^2}[\delta C_{UiUj}] 
  = \frac{4\kappa^2 V^2}{f^2} [\delta C_{ViVj}]
  =\frac{k^2(F^t_t-F^r_r)}{2f}\SHB_{ij} \notag\\
&& \qquad + \insbra{-r^2F''-nr F' +\inpare{\frac{k^2}{2n}
    -\frac{n-1}{2(n+1)}r^2\Psi}\frac{F^r_r-F^t_t}{f}}
    \gamma_{ij}\SHB, \\
&& [\delta C_{UiVj}]
 =\frac{f}{4\kappa^2 UV}\Big[
  \Big\{-r^2f F'' -(r^2f)' F' \notag\\
&&\qquad
  -\inpare{\frac{2(n-1)}{n+1}r^2\Psi-2(n-2)\lambda r^2+\frac{n-2}{n}k^2}F
  \notag\\
&&\qquad
 +\frac{r}{2}[f(F^r_r-F^t_t)]'\Big\}\gamma_{ij}\SHB
 +(n-2)k^2 F \SHB_{ij}\Big],\\
&& [\delta C_{Uijk}] =\frac{f}{2\kappa U}[\delta C_{rijk}],\quad
   [\delta C_{Vijk}] =\frac{f}{2\kappa V}[\delta C_{rijk}],
}
with $[\delta C_{ijkl}]$ as above.


\end{document}